\documentclass[a4paper,12pt]{article}
\usepackage[usenames,dvipsnames]{pstricks}
\usepackage[utf8]{inputenc}
\pdfoutput=1
\usepackage{lmodern} 
\usepackage[T1]{fontenc}
\usepackage{multicol}
\usepackage{epsfig}
\usepackage[fleqn]{amsmath}
\usepackage{amsmath,amsfonts}
\usepackage{textcomp}
\usepackage{graphicx}
\usepackage{subcaption}
\usepackage{authblk}
\usepackage[font={small,it}]{caption}
\usepackage{epstopdf}
\usepackage[pdfencoding=auto, psdextra]{hyperref}
\usepackage{bookmark}
\usepackage{float}
\hypersetup{pdfauthor={some author},pdftitle={cp paper}}
\usepackage {ifpdf}
\usepackage[margin=0.9 in]{geometry}
\usepackage[english]{babel}

\usepackage[numbers,sort&compress]{natbib}

\title{\textbf{CP-Violation phase analysis via non-trivial correlation of quarks and leptons in 3+1 scenario}}
\author{\small{Gazal Sharma\footnote{gazzal.sharma555@gmail.com}}}

\author{\small{B. C. Chauhan\footnote{chauhan@associates.iucaa.in}}}

\affil{\textit{Department of Physics and Astronomical Science, }\\
\textit{ School of Physical and Material Sciences,}\\
\textit{ Central University of Himachal Pradesh (CUHP),}\\
\textit{Dharamshala, Kangra (HP),
India 176215}}
\date{}

\begin{document}

\maketitle

\begin{abstract}
The existence and mysterious nature of sterile neutrinos are revolutionizing physics from the particle level to the cosmological scales. The recent results from the  MiniBooNE experiment at Fermi-lab observed far more $\nu_{e}$ appearance than expected, which have provided a hint about the possible existence of \textit{sterile neutrinos}. The results, if confirmed in future experiments, will have significant implications for cosmology and astroparticle physics. This will require new neutrino mass models to accommodate these additional degrees of freedom. In respect to that, the present work is just an extension of our recent work towards the CP phase analysis of Quark-lepton complementarity(QLC) model in a 3+1 scenario. The parametrization of $CKM_{4}$ and  $PMNS_{4}$ using Monte Carlo Simulation is used to estimate the texture of non-trivial correlation matrix ($V_{c_{4}}$). As such, we have successfully investigated the constrained values for sterile neutrino parameters, and also predicted the values for Dirac CP-Violation phase and the CP re-phasing invariant (J). The results obtained are consistent with the data available from various experiments, like  No$\nu$A, MINOS, SuperK and IceCube-DeepCore. Furthermore, this analysis would be very important in view of growing sterile neutrino experiments.
\end{abstract}


\section{Introduction}
The past several years in the history of neutrinos have provided us with the
experimental confirmation of three-neutrino(electron, muon and tau neutrinos) oscillations. Various other oscillation-like anomalous experimental results have lead us to the revelation of different mysteries of neutrinos, but leaving behind some issues and interpretations that require new
physics beyond Standard Model(SM). During investigating global data fit results from various experiments, so far we have got a picture which suggests that the $U_{PMNS}$ matrix contains two large and a small mixing angles; i.e. the $\theta_{23}^{PMNS}$ $\approx$ $45^\circ$, the $\theta_{12}^{PMNS}$ $\approx$ $34^\circ$ and the $\theta_{13}^{PMNS}$ $\approx$ 
$9^{\circ}$. These results when compared with the quark flavor mixing matrix i.e $U_{CKM}$ parameters, which are the ones quite settled long ago with three mixing angles that are small i.e. $\theta_{12}^{CKM}\approx 13^\circ$, $\theta_{23}^{CKM}\approx 2.4^\circ$ and $\theta_{13}^{CKM}\approx 0.2^\circ$, a certain disparity between quark and lepton mixing angles has been noticed. Since the quarks and leptons are fundamental constituents of matter and SM, the complementarity between the two is assumed to be as a consequence of a symmetry at some high energy scale. This complementarity popularly termed as `Quark-Lepton Complementarity'(QLC), that has been already explored by several authors and the possible consequences have been widely investigated in the literature \cite{Georgi:1979df,10,11,12,13}. In particular, a simple correspondence between the $U_{PMNS}$ and $U_{CKM}$ matrices have been proposed and used by several authors \cite{Ferrandis:2004mq,16,17,18}, and analysed in terms of a correlation matrix $V_{c}$ \cite{81,Chauhan:2006im,Xing:2005ur,20}.
One possible new physics interpretation mentioned earlier is about the existence of the fourth generation of neutrinos i.e. light sterile neutrinos. Those are new neutrino states which are assumed to have no weak interactions and masses of order $0.1$ - $10$ eV. 
There are bounds on the active-sterile mixing, but so far, there is no bound on the number of
sterile neutrinos and on their mass scales. Therefore, the existence of sterile neutrinos is investigated at different mass scales by various experiments; LSND, MiniBooNE, MINOS, Daya Bay, IceCube etc. The strong indication of gauge coupling unification at high scale helped us to put constraints on Dirac CP-Violation phase and CP-Violation re-phasing invariant $J$.

Despite the various successful results from the solar, atmospheric, reactor and accelerator neutrino experiments, there are experimental anomalies that cannot be explained within the standard three-neutrino framework. In particular, the possible presence of sterile neutrinos points towards the physics beyond Standard Model. So, after the successful results obtained in our previous papers for three neutrinos, here we have tried to extend our model and perform the analysis in the $3+1$ scenario. The motivating factor that lead us towards the extension of our model in 3+1 scenario was that the results obtained in our previous works \cite{81,Chauhan:2006im,Xing:2005ur,20,22}(\textit{and references therein}), are quite consistent with the recent results from No$\nu$A and IceCube \cite{19,ic}, which gives us new ray of hopes in favour of our model and its stability.
In this investigation, the QLC constrained the values of Dirac CP-Violation phase $\phi$ and the CP re-phasing invariant $J$. We have also predicted the values for two sterile neutrino mixing angles i.e. $\theta_{24}$ and $\theta_{34}$.
 
The various parts of this work are organized as follows: in section ({\bf\ref{sec:2}}), we describe in brief the theoretical framework of the QLC model in $3+1$ scenario. For the generation of $4 \times 4$ $V_{c4}$ matrix standard parametrizations were taken for $U_{CKM_4}$ and $U_{PMNS_4}$, everything is discussed in section ({\bf\ref{sec:2}}). In section ({\bf\ref{sec:3}}) results are discussed and concluded through various plots (contours, histograms, scatter..) depicting the behaviour of various neutrino parameters. Finally, the conclusions are summarised in section ({\bf\ref{sec:4}}).

\section{Phenomenology of QLC Model in 3+1 scenario}\label{sec:2}

As stated above, we observe a pattern of mixing angles of quarks and
leptons and combine them with the pursuit for unification i.e. symmetry at some high energy
leads the concept of quark-lepton complementarity i.e., QLC.
A simple correspondence between the PMNS and CKM matrices have been proposed and analyzed in terms of a correlation matrix  $V_{c}$. In our model, we proposed
 \begin{equation} \label{eq}
  V_{c}= U_{CKM} \cdot U_{PMNS} \hspace{.5cm} \rightarrow \hspace{.5cm} V_{c}= U_{CKM} \cdot \psi \cdot U_{PMNS}
 \end{equation}
\vspace{0.3cm}
\noindent where $V_{c}$ is the correlation matrix defined as a product of $U_{PMNS}$ and $U_{CKM}$.

For $3+1$ scenario a light sterile neutrino is combined with the active neutrino. In that case, we define $\psi_4$ in place of $\psi$ in equation ({\bf \ref{eq}}). So that with $\psi_4$ the equation takes the form
\begin{equation} \label{vc}
V_{c_{4}}=U_{CKM_{4}} \cdot \psi_{4} \cdot U_{PMNS_{4}}
\end{equation}

 Here the quantity $\psi_4$ is a diagonal matrix $\psi_{4}= diag(e^{\iota \psi{_i}})$ and the four phases of $\psi_i$ are set as free parameters because they are not restricted by present experimental evidences. The $3+1$ active-sterile mixing scheme is a perturbation of the standard three-neutrino mixing in which the $3 \times 3$ unitary mixing matrix U is extended to a $4 \times 4$ unitary mixing matrix, which leads to the generation of $U_{PMNS_{4}}$ lepton mixing matrix. However, the addition of a fourth generation to the standard model leads to a $4 \times 4$ quark mixing
matrix $U_{CKM_{4}}$, which is an extension of the Cabibbo-Kobayashi-Maskawa (CKM)
quark mixing matrix in the standard model.

\subsection{ \texorpdfstring{$CKM_{4}$}{CKM\_{4}} and \texorpdfstring{$PMNS_{4}$}{PMNS\_{4}} formulation}

In order to estimate the texture of $V_{c_{4}}$ we have used the $U_{CKM_{4}}$ and $U_{PMNS_{4}}$ taking reference from several works \cite{Alok,24,25,26}. The CKM matrix in SM is a $3 \times 3$ unitary matrix while in the $SM_{4}$ (this is the simplest extension of the SM, and retains all of its essential features: it obeys all the SM  symmetries and does not introduce any new ones), the $U_{CKM_{4}}$ matrix is $4 \times 4$ which can be written as

 \[
U_{CKM_{4}} =
\begin{bmatrix}
\tilde{V_{ud}} & \tilde{V_{us}} & \tilde{V_{ub}} & \tilde{V_{u{b}\prime}}\\
\tilde{V_{cd}} & \tilde{V_{cs}} & \tilde{V_{cb}} & \tilde{V_{c{b}\prime}}\\
\tilde{V_{td}} & \tilde{V_{ts}} & \tilde{V_{tb}} & \tilde{V_{t{b}\prime}}\\
\tilde{V_{{t}\prime d}} & \tilde{V_{{t}\prime s}} & \tilde{V_{{t}\prime b}} & \tilde{V_{{t}\prime{b}\prime}}
\end{bmatrix}
.
\]

Here all the elements of the matrix have their usual meanings except for ${b}\prime$ and ${t}\prime$.  Although the fourth generation quarks are too heavy to produce in LHC, but they may affect the low energy measurements, such as the quark ${t}\prime$ would contribute to $b \rightarrow s$ and $b \rightarrow d$ transitions, while the quark ${b}^{\prime}$ would contribute similarly to $c \rightarrow u$ and  $t \rightarrow c$. \\
Here, we use the Dighe-Kim (DK) parametrization defines \cite{Alok,24,25,26}.

\noindent $\tilde{V_{ud}}=1-\frac{\lambda^{2}}{2},
\tilde{V_{us}}=\lambda,
\tilde{V_{ub}}=A \lambda^{3} C e^{\iota \delta_{ub}},\\
\tilde{V_{u{b}\prime}}=p\lambda^{3}e^{-\iota\delta_{u{b}\prime}},
\tilde{V_{cd}}=-\lambda,
\tilde{V_{cs}}=1-\frac{\lambda^{2}}{2},\\
\tilde{V_{cb}}=A \lambda,
\tilde{V_{c{b}\prime}}=q\lambda^{2}e^{-\iota\delta_{c{b}\prime}},\\
\tilde{V_{td}}=A \lambda^{3}(1- C e^{\iota \delta_{ub}})+r\lambda^{4}(q e^{-\iota\delta_{c{b}\prime}}-p e^{-\iota\delta_{u{b}\prime}}),\\
\tilde{V_{ts}}=-A\lambda^{2}-q r\lambda^{3}e^{-\iota\delta_{c{b}\prime}} +\frac{A}{2}\lambda^{4}(1+r^{2}C e^{\iota \delta_{ub}}),
\tilde{V_{tb}}=1- \frac{r^{2} \lambda^{2}}{2},\\
\tilde{V_{t{b}\prime}}=r\lambda,
\tilde{V_{{t}\prime d}}=\lambda^{3}(q e_{\iota\delta_{c{b}\prime}})+ Ar\lambda^{4}(1+Ce^{\iota \delta_{ub}}),\\
\tilde{V_{{t}\prime s}}=q\lambda^{2}e^{-\iota\delta_{u{b}\prime}}+ Ar\lambda^{3}+\lambda^{4}(-p e^{-\iota\delta_{u{b}\prime}}+\frac{q}{2}e_{\iota\delta_{c{b}\prime}}+ \frac{qr^{2}}{2} e_{\iota\delta_{c{b}\prime}} ),\\
\tilde{V_{{t}\prime b}}=-r\lambda \qquad and \qquad
\tilde{V_{{t}\prime{b}\prime}}=1- \frac{r^{2}\lambda^{2}}{2}$ ,

In the presence of the sterile neutrino $\nu_{s}$, the flavor
($\nu_{\alpha}, \alpha = e, \mu, \tau, s$) and the mass eigenstates ($\nu_{i}, i =
1, 2, 3, 4$) are connected through a $4 \times 4$ unitary mixing
matrix $U_{PMNS_{4}}$, which depends on six complex parameters (i.e. it contains 6 mixing angles, and 3 oscillation-relevant
CP-violating phases) \cite{para}. In the case of one sterile neutrino, U is typically parametrized by

\begin{center}
 $
U= \tilde{R_{34}}R_{24}\tilde{R_{14}}R_{23}\tilde{R_{13}}R_{12}.$
\end{center}

Here the rotation matrices R, $\tilde R$ can be further parametrized as(for example $R_{24}$ and $\tilde R_{34}$) 
 \[
R_{24} =
\begin{bmatrix}
1 & 0 & 0&0\\
0 & 1 &0&0\\
0  &C_{24}&1&S_{24}\\
0 &-S_{24}& 0&C_{24}
\end{bmatrix}
,
\]
 \[
\tilde {R_{34}} =
\begin{bmatrix}
1 & 0& 0&0\\
0 & 1 &0 &0\\
0 & 0& C_{34}& S_{34}e^{\iota \phi}\\
0 & 0&-  S_{34}e^{\iota \phi}& C_{34}
\end{bmatrix}
,
\]
\noindent where $C_{ij}\equiv \cos\theta_{ij}$, $S_{ij}\equiv \sin\theta_{ij}$ and $\tilde S_{ij}\equiv S_{ij} e^{-\iota \phi_{ij}}$ and here,
$\phi_{ij}$ are the lepton Dirac CP phases. These phases are generalised as $\phi$, and being unconstrained we have used the same range of spread for all [$0-2\pi$] with flat distribution. Since, neutrinos are neutral particles, they can be Majorana (with neutrinos and antineutrinos being the same particle), or Dirac particles (with neutrinos and antineutrinos being different objects). Therefore, in $3+1$ scenario, there are three more CP violating phases if neutrinos are Majorana particles.
As Majorana phases do not appear in the neutrino oscillation probability, they are not measurable in the oscillation experiments.

\section{ Results and Discussion}\label{sec:3}

A non-trivial correlation between $CKM_4$ and $PMNS_4$ mixing
matrices is obtained by taking into account the phase mismatch between quark and
lepton sectors as $\psi_{4}$, a diagonal matrix $\psi_{4}= diag(e^{\iota \psi{_i}})$. 
In the $U_{CKM_{4}}$ matrix described by Dighe-Kim (DK) parametrization all the elements of $U_{CKM}$ are unitary up to ${\cal O}(\lambda^4)$. 

 The reference values for $\theta_{14}$, $\theta_{24}$ and $\theta_{34}$  are assumed to vary freely between ($0- \pi/4$). The reason behind this specific limit ($0- \pi/4$) is that all the values obtained using our reference experiments i.e. No$\nu$A, MINOS, SuperK, and IceCube-DeepCore vary between this similar range so instead of taking a specific value we have taken the whole range in our model.
 After performing the Monte Carlo simulations we estimated the texture of the correlation matrix ($V_{c_{4}}$) for two different values of $m_{{t}\prime}= $ $400 GeV$ $\&$ $600 GeV$ (where $m_{{t}\prime}$ is the mass of ${t}\prime$). We implemented the same matrix in our inverse equation and obtained the constrained results for the sterile neutrino parameters.
 Thereby we have obtained predictions for  CP-Violating re-phasing invariant and $\theta_{24}$ and $\theta_{34}$, then compared our results with the current experimental bounds are given by No$\nu$A, MINOS, SuperK, and IceCube-DeepCore experiments \cite{experi,30,31,32,33}. It has been found that the entire analysis is very less sensitive to $\theta_{14}$, which is constrained to be small by reactor experiments \cite{34}. Also in the QLC model analysis  the value obtained, is lesser as compared to the other two sets i.e. $\theta_{24}$ and $\theta_{34}$.

As we have divided our results into two parts i.e. for $m_{{t}\prime}= 400 GeV$ and $m_{{t}\prime}= 600 GeV$. The tables ~({\bf\ref{label2}}) and ~({\bf\ref{label}}) below shows the comparison of upper limits obtained using model with the four different experimental results.

\begin{table}[!hbt]
\centering
\begin{tabular}{|c|c|c|c|c|c|}
\hline 
 \textbf{Parameters} & \textbf{$\theta_{24}^{PMNS_{4}}$} & \textbf{$\theta_{34}^{PMNS_{4}}$}  \\ 
 \hline \hline
For $m_{{t}\prime} = 400 GeV$  & $6.57^{\circ}-23.36^{\circ}$ & $1.53^{\circ}-31.59^{\circ}$  \\
\hline 
For $m_{{t}\prime} = 600 GeV$ & $6.87^{\circ}-23.15^{\circ}$ & $3.78^{\circ}-32.40^{\circ}$ \\
 \hline
 \textbf{Parameters} &  $\mid U_{\mu 4}\mid^{2}$ & $\mid {U_{\tau 4}}\mid^{2}$\\
 \hline
 For $m_{{t}\prime} = 400 GeV$  &$0.0003-0.0300$ & $ 0.00-0.2031$\\
  \hline
 For $m_{{t}\prime} = 600 GeV$  & $ 0.0001-0.0236$  & $0.00-0.1432$\\
  \hline
\end{tabular}
\caption{The limits obtained on sterile mixing angles.\label{label2}}
\end{table}

\begin{table}[!hbt]
\centering
\begin{tabular}{|c|c|c|c|c|c|}
\hline 
 \textbf{Experiment} & \textbf{$\theta_{24}^{PMNS_{4}}$} & \textbf{$\theta_{34}^{PMNS_{4}}$} & $\mid U_{\mu 4}\mid^{2}$ & $\mid {U_{\tau 4}}\mid^{2}$ \\ 
 \hline \hline
No$\nu$A & 20.8 & 31.2 & 0.126 & 0.268   \\
\hline 
MINOS & 7.3 & 26.6 &0.016 & 0.20  \\
 \hline
SuperK & 11.7 & 25.1 &0.041 & 0.18   \\ 
\hline
IceCube-DeepCore & 19.4 & 22.8 & 0.11 & 0.15\\
 \hline
  \textbf{QLC Model} & \textbf{$\theta_{24}^{PMNS_{4}}$} & \textbf{$\theta_{34}^{PMNS_{4}}$} & $\mid U_{\mu 4}\mid^{2}$ & $\mid {U_{\tau 4}}\mid^{2}$ \\ 
 \hline
QLC(400 GeV) &23.36  & 31.59 &0.030 & 0.203 \\
 \hline
QLC(600 GeV) & 23.15 & 32.40 & 0.024 & 0.143\\
\hline

\end{tabular}
\caption{The upper limits of sterile mixing
parameters obtained from model(QLC) and from NO$\nu$A, MINOS, Super-Kamiokande and IceCube-
DeepCore.\label{label}}
\end{table}

\newpage
Using the same Monte Carlo procedure the absolute values of the CP re-phasing invariant 
J and CP phase(Dirac Phase) obtained are as under
\begin{center}
 $J = -0.055 - 0.055 $\\
$\mid J \mid= 0.001 - 0.055$\\
$\mid J \mid$ (Best fit)$ = 0.021$
\end{center}

\begin{center}
$
\phi= 224.64^{\circ} - 279.85^{\circ} $\\
$\phi$ (Best fit) $= 267.62^{\circ} $
\end{center}

One can see the above values predicted by the model are in very good agreement with the experimental values. In this section, we have shown our results using various plots.

\begin{figure}[!hbt]
\minipage[t]{0.35\linewidth}
 
\includegraphics[width=\textwidth]{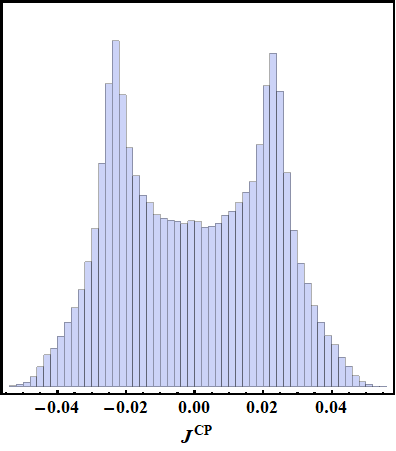}
\endminipage\hfill
\minipage[t]{0.35\linewidth}

\includegraphics[width=\textwidth]{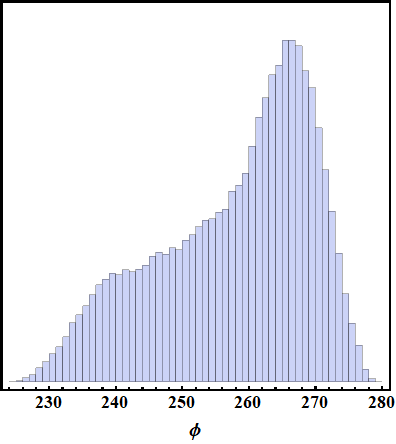}
\endminipage\hfill

\caption{ Histograms of Jarlskog re-phasing invariant (J) and Dirac CP phase ($\phi$).}
\label{1f}
\end{figure}

\newpage
In figure ({\bf\ref{1f}}) we have shown two histograms for varying range of  Jarlskog invariant J and Dirac CP phase $\phi$ on left and right panels, respectively. In the left panel we have obtained a bimodal histogram(histogram with two peaks). This implies that the range obtained for J has two values that appears most often in the data. Here one can completely see that the maximum number of J values lies near the values $\sim -0.022$ and $0.024$, which is precisely our best fit values and is comparable to the recent particle data group value range \cite{pdg}. While in the histogram on the right panel, we have obtained a distribution which is skewed left. In such plots, most of the data is clustered around the larger values and as we move towards the smaller values the number of data goes on decreasing. So, we can clearly state from the plot that the maximum number of values are gathered around $\phi= 267.62^{\circ} $ which is our best value and lies very close to the bounds given by global data analysis \cite{pdg}.

\begin{figure}[!hbt]
\minipage[t]{0.4\linewidth}
 
\includegraphics[width=\textwidth]{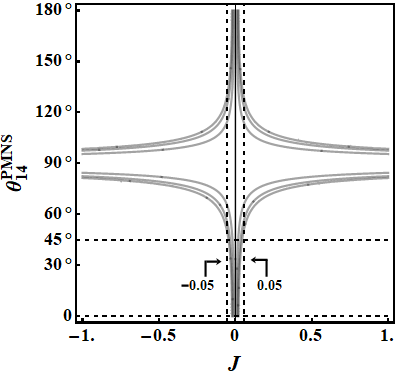}
\endminipage\hfill
\minipage[t]{0.4\linewidth}

\includegraphics[width=\textwidth]{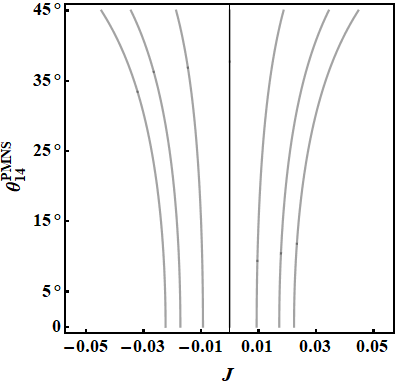}
\endminipage\hfill

\caption{ Contour plots showing the dependence of Jarlskog invariant J, $\theta_{14}^{PMNS}$ and $\theta_{23}^{PMNS}$.}
 \label{2}
\end{figure}

The contour plots ({\bf\ref{2}}) showing the correlation and the dependence of various J values on the different value ranges of $\theta_{14}^{PMNS}$ and $\theta_{23}^{PMNS}$. One can surely say that the maximum density of the value range lies between the $J=0.025-0.035$ for $\theta_{14}^{PMNS}=0^\circ- 45^\circ $ and $\theta_{23}^{PMNS}= 38.2^\circ- 53.3^\circ$. In these contour plots ({\bf\ref{2}}), the dotted lines in figure {\bf\ref{2}} (left) represents the values obtained through our analysis which we have zoomed in figure {\bf\ref{2}} (right). The 3 grey contour lines shows the various $\theta_{23}^{PMNS}$ values obtained randomly for the varying range of J and $\theta_{14}^{PMNS}$.

\begin{figure}[!hbt]
\minipage[t]{0.36\linewidth}
 
\includegraphics[width=\textwidth]{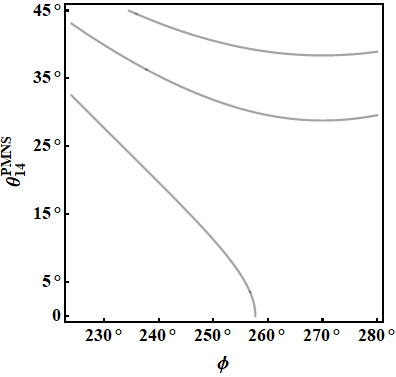}
\endminipage\hfill
\vspace{0.1cm}
\minipage[t]{0.36\linewidth}

\includegraphics[width=\textwidth]{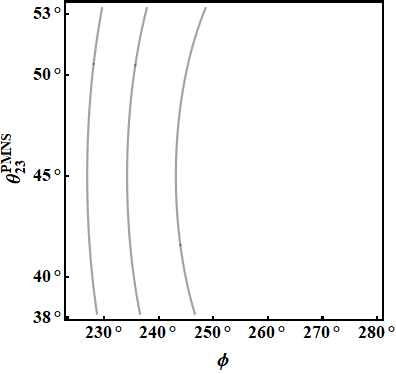}
\endminipage\hfill

\caption{ Contour plots showing the dependence of Jarlskog invariant J, $\phi$ and neutrino mixing parameters( $\theta_{14}^{PMNS}$(Left) and $\theta_{23}^{PMNS}$(Right)).}
 \label{phi2}
\end{figure}
\newpage
In figure ({\bf\ref{phi2}}), we have shown the correlation between the different values of J, $\phi$ and neutrino mixing parameters($\theta_{14}^{PMNS}$(Left) and $\theta_{23}^{PMNS}$(Right)). In the plot the dark grey lines represents the different values of $\theta_{14}^{PMNS}$(Left) and $\theta_{23}^{PMNS}$(Right) obtained randomly.

\begin{figure*}[!hbt] 
\minipage[t]{0.35\linewidth}
 
\includegraphics[width=\textwidth]{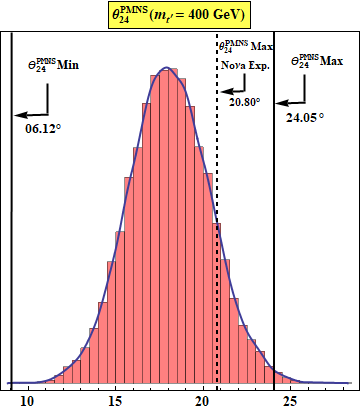}
\endminipage\hfill
\vspace{0.08cm}
\minipage[t]{0.35\linewidth}
 
\includegraphics[width=\textwidth]{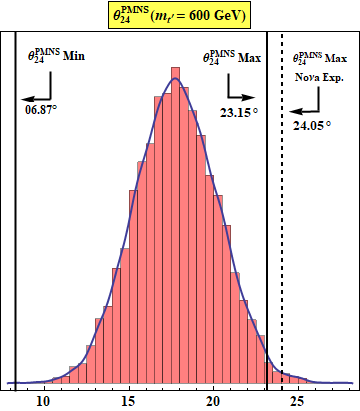}
 \endminipage\hfill
\caption{ Probability density distribution of $\theta_{24}^{PMNS}$ for 
$m_{{t}\prime}= 400 GeV$(Left) and $m_{{t}\prime}= 600 GeV$(Right) \cite{22}.}
 \label{3}
\end{figure*}

\begin{figure*}[!hbt] 
\minipage[t]{0.35\linewidth}
 
\includegraphics[width=\textwidth]{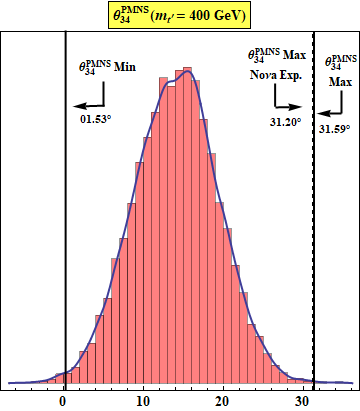}
\endminipage\hfill
\vspace{0.08cm}
\minipage[t]{0.35\linewidth}
 \includegraphics[width=\textwidth]{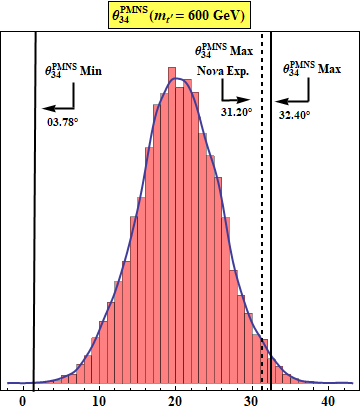}
 \endminipage\hfill
\caption{ Probability density distribution of $\theta_{34}^{PMNS}$ for 
$m_{{t}\prime}= 400 GeV$(Left) and $m_{{t}\prime}= 600 GeV$(Right) \cite{22}.}
 \label{4}
\end{figure*}

The results are divided varying upon the value $m_{{t}\prime}= $ $400 GeV$ $\&$ $600 GeV$.
We obtain histograms of the probability density 
functions for  $\theta_{24}^{PMNS}$ and $\theta_{34}^{PMNS}$ for both $400 GeV$ and $600 GeV$, respectively and show a comparison between the two. In figure ({\bf\ref{3}}) and ({\bf\ref{4}}), we have compared our results for the upper values from No$\nu$A experiment. Here the dashed line is for $\theta_{24}^{PMNS}$ and $\theta_{34}^{PMNS}$ experimental the thick solid lines are the 3-$\sigma$ upper and lower values of $\theta_{24}^{PMNS}$ and $\theta_{34}^{PMNS}$ obtained using the QLC model. In figure ({\bf\ref{3}}), we have shown this quite nicely the left panel of the figure is for
$\theta_{24}^{PMNS}$ for $m_{{t}\prime}= 400 GeV$, whereas the
right panel shows the $\theta_{24}^{PMNS}$ for $m_{{t}\prime}= 600 GeV$ and same is in figure ({\bf\ref{4}}) for varying range of $\theta_{34}^{PMNS}$ .

\begin{figure}[!hbt]
\minipage[t]{0.4\linewidth}
 
\includegraphics[width=\textwidth]{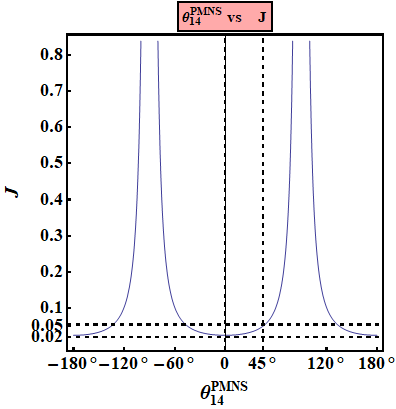}
\endminipage\hfill
\minipage[t]{0.4\linewidth}
 
\includegraphics[width=\textwidth]{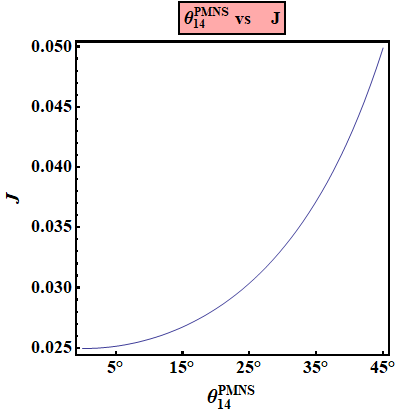}
\endminipage\hfill

\caption{Correlation plot between J and $\theta_{14}^{PMNS}$.}
 \label{5}
\end{figure}

\begin{figure}[!hbt]
\minipage[t]{0.4\linewidth}
 
\includegraphics[width=\textwidth]{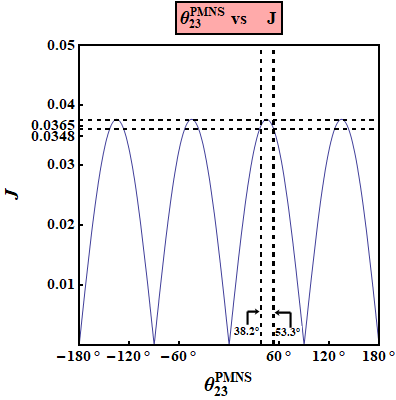}
\endminipage\hfill
\minipage[t]{0.4\linewidth}
 
\includegraphics[width=\textwidth]{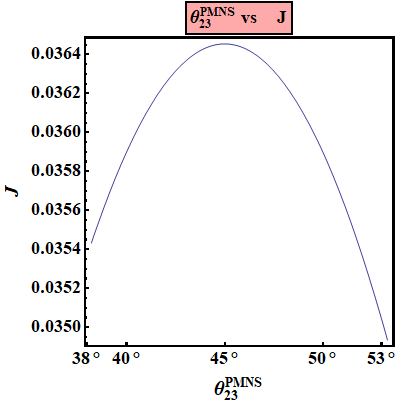}
 \endminipage\hfill
\caption{Correlation plot between J and $\theta_{23}^{PMNS}$.}
 \label{6}
\end{figure}
\newpage
The plots in figure ({\bf\ref{5}} and {\bf\ref{6}}), we have shown the correlation between the Jarlskog invariant $J$ with
$\theta_{14}^{PMNS}$ and $\theta_{23}^{PMNS}$, respectively. Here, in these two also the dotted lines in the figure in the left panel represents the values obtained through our analysis . In both figures, the plots on right panel are the zoomed portion from the full range obtained through QLC model calculations.

\section{Conclusions}\label{sec:4}
It is very hard to understand these type of relations in ordinary bottom-up
approaches where the quarks and leptons are treated separately with no specific connections
between them especially about its further extension in the fourth generation. One might ask or question about the stability of the framework that we have used in order to carry out our entire analysis which is the extension of SM in fourth generation. And one might argue that the four generation scenarios are strongly disfavoured. This is true, unless a substantial modification is realized for the scalar sector. However, such an extension of the Standard Model (with massive neutrinos) is excluded by several authors, e.g. see references \cite{x,y}. As such, during the starting period of the discovery of Higgs particle, data was not so precise that the possibility of the fourth generation coupling to the Standard Model Higgs doublet was not introduced in a different Beyond Standard Model(BSM) scenarios. But, such options were ruled out as the LHC data develops gradually, which let many scenarios to go beyond SM i.e. BSM. It has been noted that such fourth generation is hidden during the single production of the Higgs Boson, while it shows up when one considers the double Higgs production i.e. $gg \rightarrow hh$ which can be considered in a different framework of a two Higgs doublet model (2HDM) \cite{x1,g,gg,y1}. This is the framework that we have taken to carry out our analysis which is well favored by the work done and published in 2018 \cite{35}. In that work, they have shown that the current Higgs data does not eliminate the possibility of a sequential fourth generation that gets the masses through the same Higgs mechanism as the first three generations.

 As far as the stability of the QLC relation is concerned, the kind of model relation obtained with some flavor symmetry has been checked by the several works and concluded that the stability under the RGE 
effects and the radiative corrections are small. We have performed numerical simulations to investigate the sterile neutrino mixing angles and the CP phase and re-phasing invariant. This might help to look forward the understanding of QLC model in a better way. The expert eye and deep knowledge with some theoretical ground can explain QLC relations precisely. We do require some strong models from the Grand Unified Theories(GUT), which sometimes also unify quarks and leptons and provide a framework to construct a proper theory in which QLC relation can be embedded in a natural way.

The consequences of the model are the predictions for CP-Violating phase
invariant $J$, Dirac phase $\phi$ and the sterile neutrino parameters.  The results obtained numerically and analytically this paper stood in a good agreement with the experimental data.  Our analysis would be very important in view of growing sterile neutrino experiments as well as the CP Phase analysis.

\section{Acknowledgements}
B.C. Chauhan acknowledges the financial support provided by the University Grants Commission(UGC), Government of India vide Grant No. UGC MRP-MAJOR-PHYS-2013-12281. We thank IUCAA for providing research facilities during the completion of this work.

\end{document}